\documentclass{WileyMSP-template}
\usepackage{amsmath}
\usepackage{mathrsfs}
\usepackage{ragged2e}
\begin{document}

\title{Experimental Demonstration of Sequential Multiparty Quantum Secret Sharing and Quantum Conference Key Agreement}
\maketitle

\author{Shuaishuai Liu}
\author{Zhengguo Lu}
\author{Pu Wang}
\author{Yan Tian}
\author{Qing Lu}
\author{Xuyang Wang}
\author{Yongmin Li*}

\begin{affiliations}
Shuaishuai Liu, Zhengguo Lu, Pu Wang, Yan Tian, Qing Lu, Xuyang Wang,  Prof. Yongmin Li\\
State Key Laboratory of Quantum Optics and Quantum Optics Devices, Institute of Opto-Electronics, Shanxi University, Taiyuan 030006, China\\
Collaborative Innovation Center of Extreme Optics, Shanxi University, Taiyuan 030006, China\\
Email Address: yongmin@sxu.edu.cn

\end{affiliations}



\begin{abstract}
\justifying
Quantum secret sharing (QSS) and quantum conference key agreement (QCKA) provide efficient encryption approaches for realizing multi-party secure communication,  which are essential components of future quantum networks. We present three practical, scalable, verifiable $(k, n)$ threshold QSS protocols that are secure against eavesdroppers and dishonest players. The proposed QSS protocols eliminate the need for each player preparing the laser source and laser phase locking of the overall players. The dealer can implement the parameter evaluation and get the secret information of each player without the cooperation from other players. We consider the practical security of the proposed QSS systems with Trojan-horse attack, untrusted source intensity fluctuating and untrusted noisy sources. Our QSS systems are versatile, they can support the QCKA protocol by only modifying the classic post-processing and requiring no changes to the underlying hardware architecture. We experimentally implement the QSS and QCKA protocol with five parties over 25 km (55 km) single mode fibers, and achieve a key rate of 0.0061 ($7.14\times10^{-4}$) bits per pulse. Our work paves the way for the practical applications of future QSS and QCKA.
\end{abstract}
\justifying
\section{Introduction}

In recent years, quantum communication has made significant breakthroughs, in particular, quantum key distribution (QKD) \cite{1,2,3} ensures secure communication between legitimate parties based on the principles of quantum mechanics. The invention of QKD provides an effective approach to solve the point-to-point security key distribution between two users. Inspired by the idea of QKD and classical cryptography protocols \cite{4,5,6}, the quantum secret sharing (QSS) \cite{7} and quantum conference key agreement (QCKA) \cite{8} using multiparticle Greenberger-Horne-Zeilinger (GHZ) entangled states were proposed. The QSS combines quantum cryptography with classical secret sharing and uses quantum state as a secret encoding carrier. The secret message is divided into $n$ pieces and distributed to $n$ players in an appropriate way \cite{7}. For a $(k,n)$ threshold protocol, if no less than $k$ players combine their pieces of information together, the secret message can be recovered \cite{4}. QSS can protect secret message from the eavesdroppers and dishonest players, and has important applications in key management, identity authentication, remote voting, and quantum sealed-bid auction. The task of QCKA is to establish a common secret key among $n$ players. All players can encrypt the public messages and decrypt the encrypted public messages broadcasted by other players, whereas the eavesdroppers cannot obtain any public messages broadcasted by the players \cite{8}.

At present, a variety of QSS and QCKA protocols have been proposed. Depending on the quantum resources employed, the discrete variable QSS including the entangled state QSS \cite{7,9,10,11,12,13}, the single qubit QSS \cite{14,15,16}, the single qudit QSS \cite{17,18,19}, and the post-selected multipartite entanglement state QSS \cite{20} have been investigated. The continuous variable QSS with the entangled state \cite{21,22,23,24} and coherent state \cite{25,26,27,28} were also presented. Furthermore, QCKA with multipartite entangled state \cite{8,29,30,31,32,33}, three party QKD \cite{34}, and measurement-device-independent (MDI) type \cite{20,35,36} have been considered.

The above works significantly improve the feasibility of QSS and QCKA. However, there are still key limitations in security and practicability. For instance, the single qubit QSS protocol is vulnerable to Trojan horse attacks \cite{37,38} where an eavesdropper sends a signal to the player’s secure station and unambiguously determine the private information by measuring the output signals. The QSS \cite{7,9,10,11,12,13} and QCKA \cite{8,29,30,31,32,33} based on the GHZ entangled state are appealing, but the scalability is a challenge due to the difficulty of generation, manipulation of multi-partite entangled states with very large dimension at present \cite{39}. The post-selection GHZ entangled state QSS and QCKA alleviate this issue \cite{20,35}. However, the implementation of this scheme requires the intervention of multiple players, which increases the complexity of the experiment. Continuous variable QSS (CV-QSS) based on coherent states has good compatibility with telecom  techniques \cite{25,26,27}. Unfortunately, current coherent-state CV-QSS protocol requires that all players have to prepare their own laser sources (although there is a QSS scheme to solve the laser source problem, it is still difficult to achieve with the current experimental technology \cite{28}), and the laser phases of all players should be strictly stabilized, which adds considerable complexity and cost to the system. On the other hand, the superposition of channel excess noises from other players severely reduces the secret key rate due to the joint measurement by the dealer \cite{25,26,28}. Furthermore, most of the existing QSS and QCKA require dedicated hardware devices and many QSS are $(n, n)$ schemes, which limit the flexibility and versatility of the protocols.

To solve above issues, in this paper we propose three practical, scalable, and verialbe $(k, n)$ threshold CV-QSS protocols: dense wavelength division multiplexing (DWDM) QSS protocol, multiple sideband modulation (MSB) QSS protocol and their composite protocol. In contrast to previous works, our protocols do not require each player preparing a laser source and the phase locking of the overall lasers, furthermore, the evaluation of the channel parameters for each player is independent, which significantly reduce the complexity of QSS system and increase the secure secret key rate and transmission distance. The proposed QSS schemes are versatile and flexible. They can switch between QSS and QCKA just by switching the classical post-processing program and no modification of hardware devices are required. We perform strict security analysis for Trojan horse attacks and the untrusted sources intensity fluctuation and noise. The protocols are proved to be secure against eavesdroppers on the quantum channel and dishonest players. We experimentally demonstrate the QSS and QCKA protocols with five-party over long-distance single mode fiber, and investigate the excess noise variations versus the number of the player and fiber length.

\section{Quantum protocols}
\subsection{\label{sec:level3}The QSS protocols}
\subsubsection{\label{sec:level4} The DWDM-QSS and MSB-QSS protocols}
\begin{figure}
	\centering
	\includegraphics[width=3.4in]{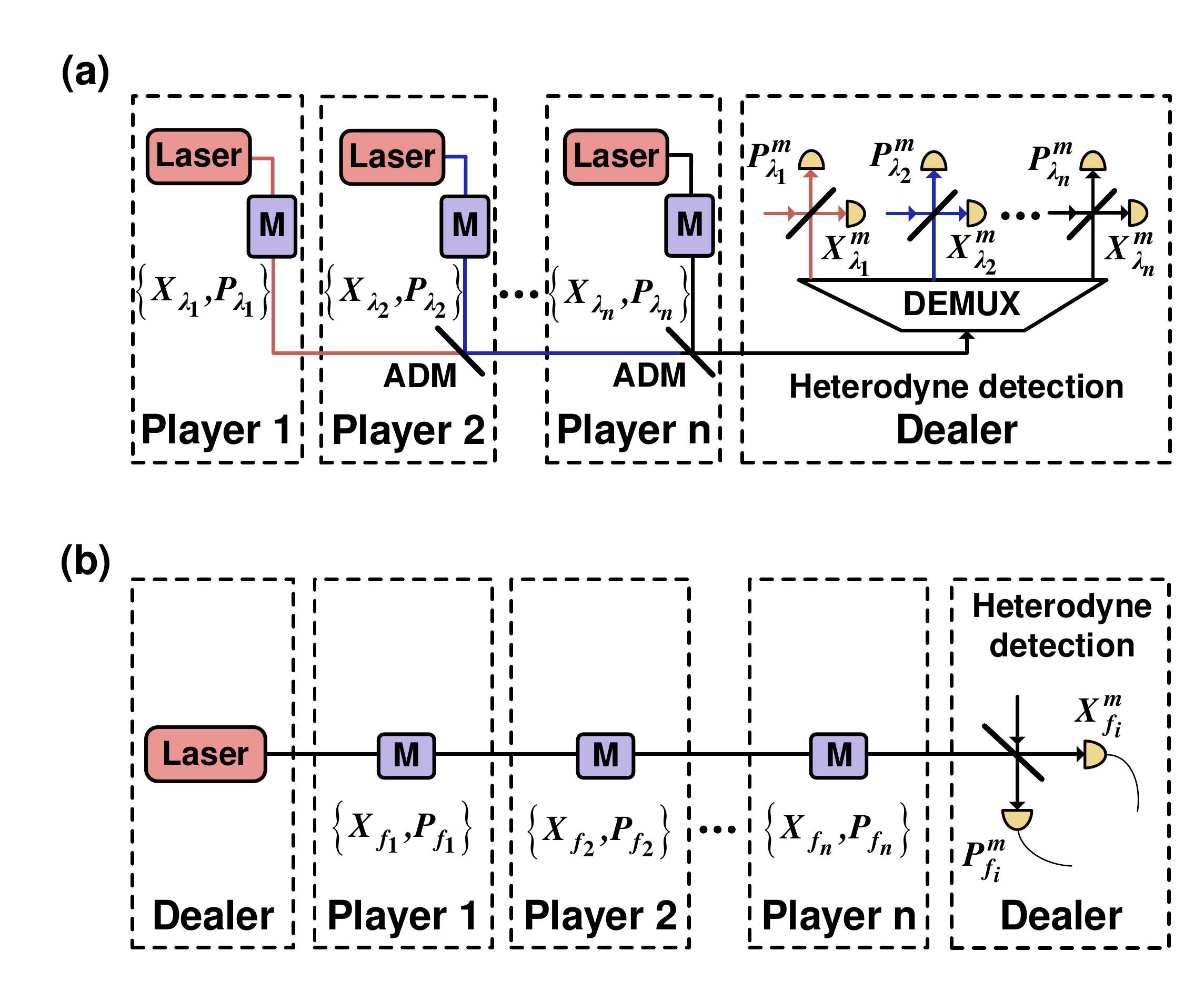}
	\caption{\label{Fig_1}The QSS protocols. M, modulator; ADM, add/drop multiplexer; DEMUX, demultiplexer. a) The DWDM-QSS scheme: each player prepares a laser source of different wavelength and implements the Gaussian modulation to encode the key information, then the signal fields with different wavelengths are multiplexed via ADM. After transmission, the dealer demultiplexs the signals of different players and measures them separately by heterodyne detection. b) The MSB-QSS scheme: the dealer prepares a coherent laser source that passes through each player in sequence, and all players implement independent Gaussian modulation to encode the information in different sidebands of the laser field. After transmission, the dealer measures the signals by heterodyne detection and extracts the corresponding sideband information in terms of the encoding rules of the players.}
\end{figure}

The sketch of the DWDM-QSS protocol is shown in Figure~\ref{Fig_1}a, each player prepares a laser source of different wavelength in the secure station and performs Gaussian modulation to encode the information. Then the signal fields with different wavelengths are multiplexed via the add/drop multiplexer (ADM) and sent to the dealer through a common quantum channel. The dealer demultiplexs the relieved signal fields via a demultiplexer (DEMUX) and measures them separately via heterodetection. The detailed steps are as follows.

Step 1. All players prepare a laser source of different wavelength ${{\lambda }_{i}}$, $i\in \left\{ 1,2,\ldots ,n \right\}$. The first player modulates the laser and prepares a coherent state $\left| {{X}_{{\lambda }_{1}}}+i{{P}_{{\lambda }_{1}}} \right\rangle $ with a pair of Gaussian random numbers $\left\{ {{X}_{{\lambda }_{1}}},{{P}_{{\lambda }_{1}}} \right\}$ and sends the prepared state to the neighboring player.

Step 2. The second player prepares the coherent state $\left| {{X}_{{\lambda }_{2}}}+i{{P}_{{\lambda }_{2}}} \right\rangle $ and adds it into the quantum channel via ADM. Above procedure is repeated until the $n$th player prepares the coherent state $\left| {{X}_{{\lambda }_{n}}}+i{{P}_{{\lambda }_{n}}} \right\rangle $ and adds it into the quantum channel.

Step 3. After the quantum states of all players reaches the dealer through a common quantum channel, the dealer uses a demultiplexer to separate the received quantum states and  measures them using heterodyne detection. The measurement results (raw data) are $\left\{ X_{{\lambda }_{1}}^{m},P_{{\lambda }_{1}}^{m} \right\},\left\{ X_{{\lambda }_{2}}^{m},P_{{\lambda }_{2}}^{m} \right\},\cdots ,\left\{ X_{{\lambda }_{n}}^{m},P_{{\lambda }_{n}}^{m} \right\}$.

Step 4. Repeat the above steps until enough raw keys are accumulated. At this stage, the quantum distribution phase of the protocol is completed.

Step 5. The dealer and each player independently evaluate the channel parameters including the quantum channel transmittance $\left\{ {{T}_{{\lambda }_{1}}},{{T}_{{\lambda }_{2}}},\cdots ,{{T}_{{\lambda }_{n}}} \right\}$ and excess noise $\left\{ {{\varepsilon }_{{\lambda }_{1}}},{{\varepsilon }_{{\lambda }_{2}}},\cdots, {{\varepsilon }_{{\lambda }_{n}}} \right\}$ by using the same procedure as that of the continuous variable QKD (CV-QKD). Based on the channel parameters, the key rates between the dealer and each player can be estimated. If all of them are positive, the dealer selects the lowest key rate $K_{\lambda _{i}}$ among all players as the key rate of the QSS, that means the QSS works at the rate of the worst performing player, then using the data reconciliation and privacy amplification, the secure keys $\left\{ {{S}_{{\lambda }_{1}}},{{S}_{{\lambda }_{2}}},\cdots, {{S}_{{\lambda }_{n}}} \right\}$ are distilled. If a $(n,n)$ threshold secret sharing is desired, it is straightforward to use the new key $S={{S}_{{\lambda }_{1}}}\oplus {{S}_{{\lambda }_{2}}}\oplus \cdots \oplus {{S}_{{\lambda }_{n}}}$ to encode the secret message $M$. 

Step 6. For a $(k, n)$ threshold QSS, the dealer randomly selects a $k-1$ power polynomial $f\left( {{S_{{\lambda }}}} \right)$ in the finite field ${{Z}}$, where $f\left( {{S}_{{\lambda }}} \right)=S+{{a}_{1}}S_{{\lambda }}^{1}+{{a}_{2}}S_{{\lambda }}^{2}+\cdots +{{a}_{k-2}}S_{{\lambda }}^{k-2}+{{a}_{k-1}}S_{{\lambda }}^{k-1} $. Here, the polynomial coefficients $\left\{S, {{a}_{1}}, {{a}_{2}},\cdots, {{a}_{k-1}} \right\}\in {{Z}}$ and $S$ is the sharing secret key. The dealer calculates $\left\{ {{S}_{{\lambda }_{i}}},f\left( {{S}_{{\lambda }_{i}}} \right) \right\}$, and selects a Hash function $H\left( {S_{{\lambda }}} \right)$ to calculate the authentication tag $\left\{ {H}\left( {{S}_{{\lambda }_{1}}} \right),{H}({{S}_{{\lambda }_{2}}}),\cdots, {H}\left( {{S}_{{\lambda }_{k}}} \right),\cdots ,{{H}}\left( {{S}_{{\lambda }_{n}}} \right) \right\}$. Next, the dealer sends $f( {{S}_{_{{\lambda }_{i}}}})$, the Hash function, and the authentication tags to each player through the authenticated classical channel.

Step 7. Each player know $f\left( {{S}_{{\lambda }_{i}}}\right)$  and the authentication tags  of all players. If $k$ players want to reconstruct the sharing secret keys, they use the Hash function to calculate the authentication tags $\left\{ H^{\prime}\left( {{S}_{{\lambda }_{1}}} \right),H^{\prime}({{S}_{{\lambda }_{2}}}),\cdots ,H^{\prime}\left( {{S}_{{\lambda }_{k}}} \right) \right\}$ and compare them with those sent by the dealer. By checking the consistency of the authentication tag, the dishonest players can be discovered. After verification, the sharing secret key $S$ can be calculated directly using the Lagrange interpolation formula:
\begin{equation}\label{1}
	S=\sum\limits_{i=1}^{k}{f\left( {{{S}_{\lambda}}_{i}} \right)\prod\limits_{l=1.l\ne i}^{k}{\frac{{{{S}_{\lambda}}_{l}}}{{{{{S}_{\lambda}}_{l}}}-{{{S}_{\lambda}}_{i}}}}}.
\end{equation}

Although above procedures are classical, we take each distributed key ${{{S}_{\lambda}}_{i}}$ as a independent variable of a polynomial $f\left( {{S}_{{\lambda }}}\right)$ and combine it with a Hash function, which makes our scheme secure against eavesdroppers and dishonest players in both the quantum distribution stage and the key reconstruction stage.

The MSB-QSS protocol is depicted in Figure~\ref{Fig_1}b, the dealer prepares a laser source that passes through each player in sequence and the players perform independent Gaussian modulation to encode the key information at different sidebands of the light field \cite{40}. The dealer extracts the encoded signals by resolving the sidebands that are modulated by the players. The detailed procedure is as follows.

Step 1. The dealer first prepares a coherent laser source. The first player prepares a coherent state $\left| {{{X}_{{f}_{1}}}+i{{P}_{{f}_{1}}}} \right\rangle $ with weak modulation at sideband frequency ${{f}_{1}}$ of the light field and sends the modulated light field to the neighboring player.

Step 2. The rest $n-1$ players prepares the coherent state  $\left| {{X}_{{f}_{i}}}+i{{P}_{{f}_{i}}} \right\rangle $, $i\in \left\{2,3,\ldots ,n \right\}$ on different sidebands respectively, and the $n$th player sends the prepared states to the dealer via the quantum channel.

Step 3. The dealer measures the received signal modes by performing heterodyne detection, and extracts the corresponding  sideband information of each player. The measurement results $\left\{ X_{{f}_{1}}^{m},X_{{f}_{1}}^{m} \right\}$,$\left\{ X_{{f}_{2}}^{m},P_{{f}_{2}}^{m} \right\}$,$\cdots$ ,$\left\{ X_{{f}_{n}}^{m},P_{{f}_{n}}^{m} \right\}$ are recorded as raw data.

Step 4. The remaining procedures are the same as those of the DWDM-QSS protocol.
\subsubsection{\label{sec:level5}The composite QSS protocol}
\begin{figure}
	\centering
	\includegraphics[width=3.4in]{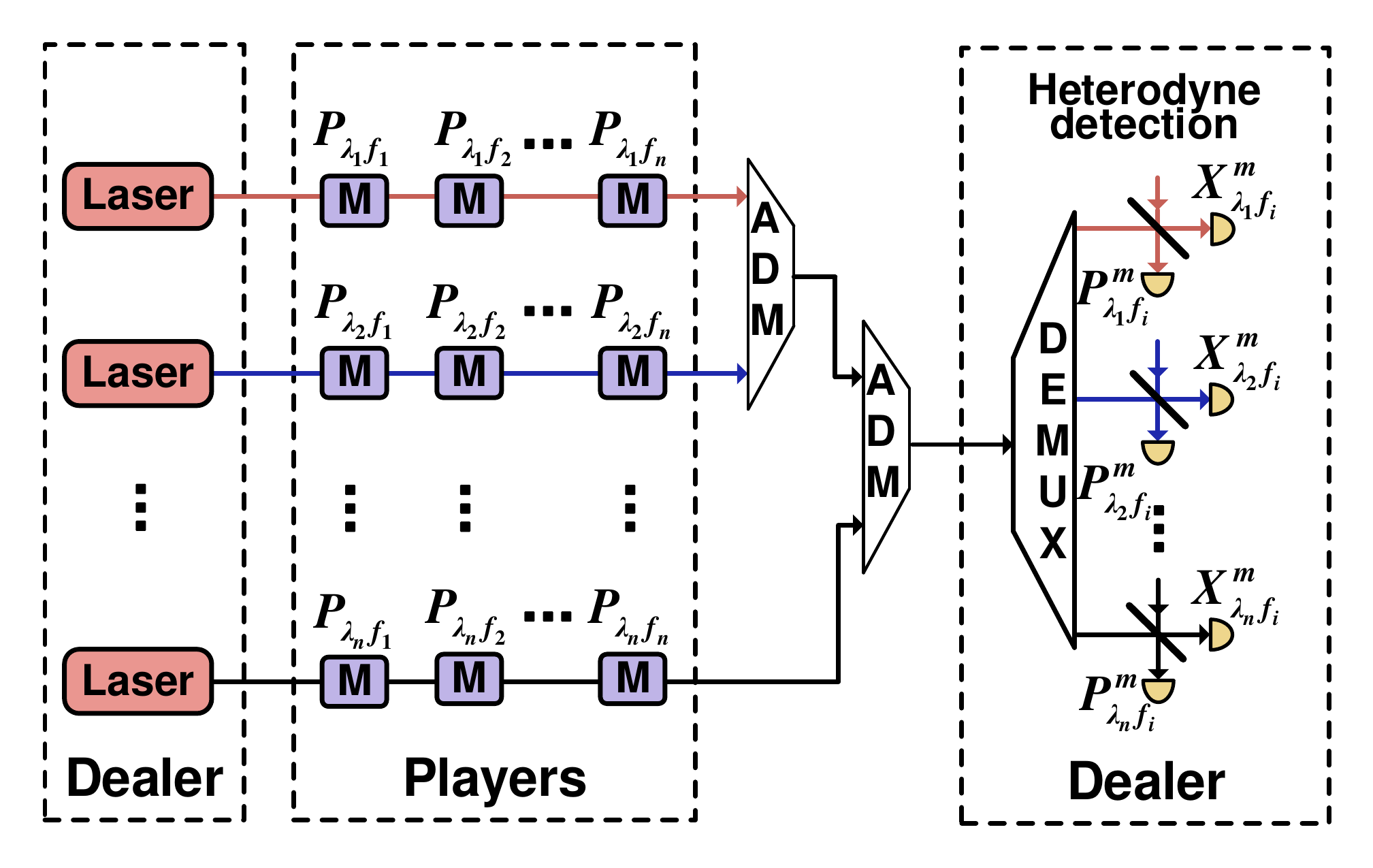}
	\caption{\label{Fig_2}The composite QSS  protocol. The dealer prepares a number of laser sources of different wavelengths and sends them to the neighboring players. The players use Gaussian modulation to encode their information in different sidebands of the lasers and subsequently send the modulated light fields to the next player. Next, all the signal fields are coupled into one fiber channel through the ADMs and sent to the dealer. Finally, the dealer demultiplexs all players's signal fields and measures them with heterodyne detection.}
\end{figure}

To enhance the versatility and practicality of the QSS, we combine the DWDM-QSS and MSB-QSS protocols to form a composite QSS scheme as illustrated in Figure~\ref{Fig_2}. In this scheme, the dealer prepares a number of laser sources of different wavelengths and sends them to adjacent players. For each laser source, the players implement independent Gaussian modulation to encode their information in different sidebands of the light field and subsequently send the modulated light field to the next one. Next, all the signal fields are coupled into the common quantum channel through the ADMs and sent to the dealer. Finally, the received signal fields are separated by the dealer using a demultiplexer and measured separately with the heterodyne detection. The detailed process is similar to that of DWDM-QSS and MSB-QSS schemes.

\subsection{\label{sec:level6}The QCKA protocol}
Considering actual application scenarios, a quantum network should not support only a single protocol. On the premise of not changing the underlying architecture, it is desired that the network can support multiple protocols which can be conveniently switched according to the needs of the players. Such a network structure is flexible and versatile \cite{20,27,35}. If a network can only implement a specific protocol with dedicated hardwares, it means that the network is not very practical.

Our experimental system is flexible and versatile and can be switched between QSS and QCKA on demand. The system can be used to implement QCKA without modifying any hardware devices, one only need to switch the corresponding post-processing procedure. Below we present the implementation process of QCKA in detail.

Step 1. By utilizing the same quantum stage as that of the QSS scheme, the dealer
establishs different quantum keys $\left\{ {{S}_{{\lambda }_{1}{f}_{1}}},{{S}_{{\lambda }_{1}{f}_{2}}} ,{{S}_{{\lambda }_{2}{f}_{1}}},{{S}_{{\lambda }_{2}{f}_{2}}},\cdots ,{{S}_{{\lambda }_{n}{f}_{n}}} \right\}$ with all players. The dealer selects the lowest secret key rate ${{K}_{{\lambda _{i}}{f_{i}}}}$ among all players, that means the QCKA works at the key rate of the worst performing player.

Step 2. The dealer prepares a common secret key ${{S_{c}}}$, which are encrypted using the player's quantum secret key ${{S}_{{\lambda }_{i}{f}_{i}}}$, ${{S}_{e}}={{S}_{{\lambda }_{i}{f}_{i}}}\oplus {{S_{c}}}$, and then sent to the designated players through the authenticated classical channel. Then the players decrypt the encrypted keys with their own quantum secret key, recover the common secret key ${{S_{c}}}={{S_{e}}}\oplus {{S}_{{\lambda }_{i}{f}_{i}}}$ \cite{30}.

Our QCKA scheme has following advantages compared with the QCKA based on multipartite GHZ entangled states. \emph{Quantum state preparation:} our scheme only requires off-the-shelf telecom components such as commercial narrow linewidth lasers, amplitude and phase modulators, thus the state preparation process is simple and low-cost. In contrast, the preparation of large scale multipartite GHZ entanglement is still a challenge with the current experimental technology. \emph{Scalability:} our scheme can be conveniently extended to plenty of players (on the order of hundreds), while the scalability is difficult for multipartite GHZ entanglement scheme due to the multiphoton coincidence counting is required for the key generation. In this case, the key rate will decline exponentially versus the total transmission loss \cite{33}.

\section{\label{sec:level7}Security analysis}
For the MSB-QSS and composite QSS scheme, similar to the theoretical framework of plug-and-play CV-QKD, the transmission of the laser source from the dealer to the players can be controlled by Eve. In this case, Eve may performs potential attacks. In this section, we analyze the security of our protocol under the condition of Trojan horse attacks, untrusted source intensity fluctuation, and untrusted source noise.
\begin{figure}[!htbp]
	\centering
	\includegraphics[width=3.4in]{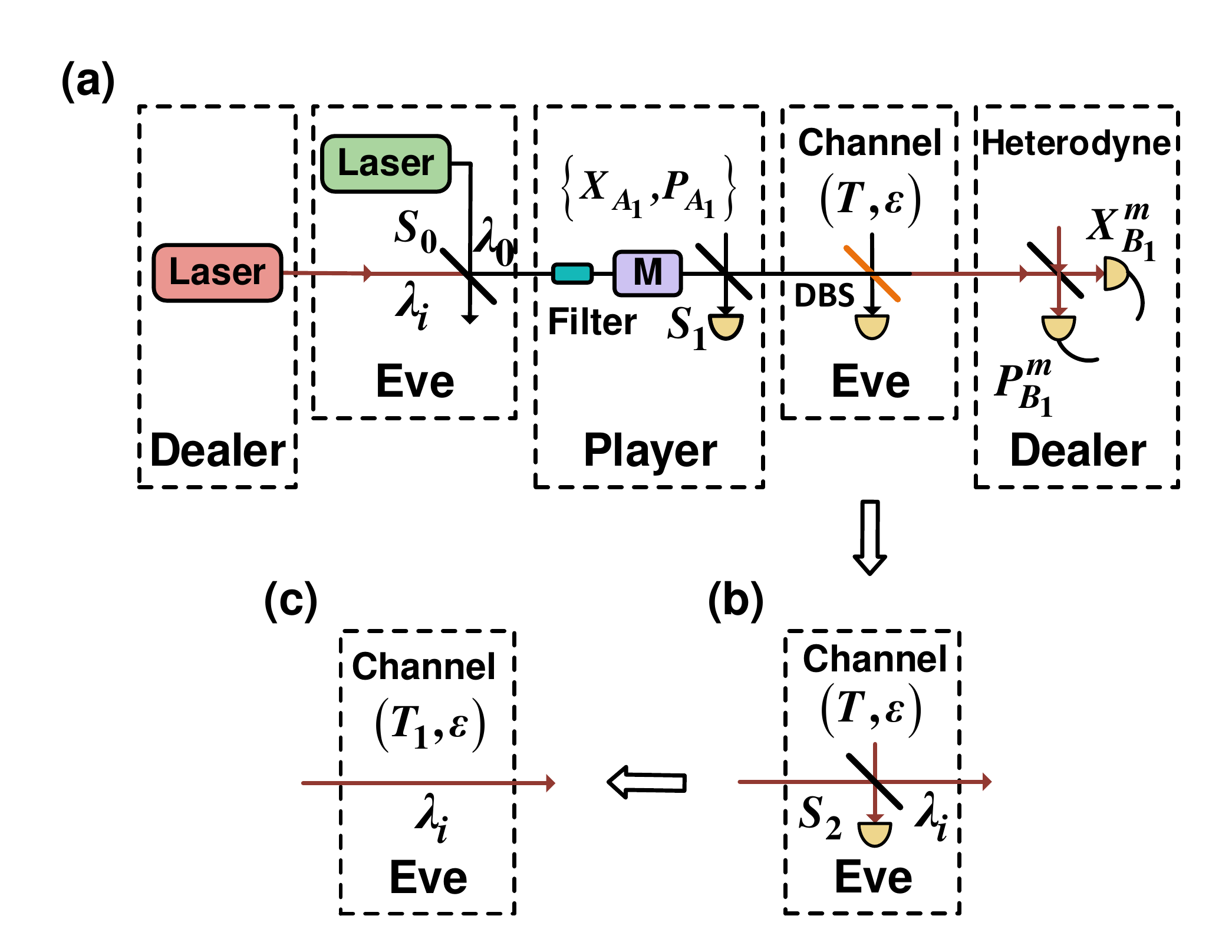}
	\caption{\label{Fig_3}Schematic diagram of the Trojan horse attack model. a) Eve uses a beam splitter to inject her probe light at wavelength of ${{\lambda }_{0}}$ into the player's station. After being modulated by the player, the probe light is separated by a DBS at outside of the station. Then Eve can acquire the key information by measuring the probe beam. b) The effect of the attack has nothing to do with the specific value of ${{\lambda }_{i}}$. We can assume ${{\lambda }_{0}}={{\lambda }_{i}}$, and replace the DBS with beam splitter $S_2$ with transmittance ${S_{2}}={I_{{\lambda}_{i}}}/{(I_{{\lambda}_{i}}+I_{{\lambda}_{0}})}$ at $I_{{\lambda}_{0}}$. c) The Trojan horse attack is equivalent to decrease the transmittance of the untrusted quantum channel from $T$ to $T_1=TS_2$.}
	
\end{figure}

\subsection{\label{sec:leve21}Security proof of Trojan horse attacks}
Due to the bidirectional feature of the MSB-QSS and composite QSS scheme, the Trojan horse attack should be considered. As shown in Figure~\ref{Fig_3}a, Eve can use a beam splitter with transmittance of ${S}_{0}$ to couple her probe laser at wavelength of ${{\lambda }_{0}}$ with the laser at wavelength of ${{\lambda }_{i}}$, $i\in \left\{ 1,2,\ldots ,n \right\}$ sent by the dealer and send them to the player. The probe laser will carry the key information after being modulated by the modulators of the players. Just at the outside of the player's station, Eve uses a dichroic beam splitter (DBS) to separate the probe laser for her measurement to obtain the key information.

To deal with this attack, we insert a 50-GHz narrow-band optical filter (0.4 nm bandwidth) into the player's input port, thereby limiting Eve's probe laser wavelength to $\left| {{\lambda }_{0}}\text{-}{{\lambda }_{i}} \right|\le 0.2$ nm. A beam splitter with transmittance of ${S}_{1}$  is added after the modulator to monitor the modulated light fields. The measured light intensity is given by
\begin{equation}\label{2}
	{{I_{m}}}={{\eta }_{{{\lambda }_{i}}}}{{I}_{{{\lambda }_{i}}}}+{{\eta }_{{{\lambda }_{0}}}}{{I}_{{{\lambda }_{0}}}}+I_{el},
\end{equation}
where ${\eta }_{{\lambda }_{i}}$ and ${\eta }_{{\lambda }_{0}}$ are the total detection efficiency of the player and Eve, respectively, including the modulator's loss, split ratio of beam splitter, and the photodiode's quantum efficiency. ${{I}_{{{\lambda }_{i}}}}$ and ${{I}_{{{\lambda }_{0}}}}$ are the light intensity of the player and Eve respectively, and ${{I}_{\text{el}}}$ is the electronic noise of the monitoring detector. Since a weak electro-optic modulation is employed, the modulation variance is proportional to the light intensity of the modulated laser. The modulation variance of the player and Eve can be expressed as
\begin{equation}\label{3}
	{{V}_{A}}={{M}_{{{\lambda }_{i}}}}{{I}_{{{\lambda }_{i}}}},\text{  }{{V}_{{{\lambda }_{0}}}}={{M}_{{{\lambda }_{0}}}}{{I}_{{{\lambda }_{0}}}},
\end{equation}
where ${{M}_{{{\lambda }_{i}}}}$ and ${{M}_{{{\lambda }_{0}}}}$ are the modulation coefficients of the electro-optic modulators. Substituting Eq. (\ref{3}) into Eq. (\ref{2}) we get
\begin{equation}\label{4}
	{{I_{m}}}=\frac{{{\eta }_{{{\lambda }_{i}}}}{{V}_{{A}}}}{{{M}_{{{\lambda }_{i}}}}}+\frac{{{\eta }_{{{\lambda }_{0}}}}{{V}_{{{\lambda }_{0}}}}}{{{M}_{{{\lambda }_{0}}}}}+I_{el}.
\end{equation}

Considering that the wavelength of the probe light is very close to the wavelength of the dealer' laser ${{\lambda }_{0}}\approx {{\lambda }_{i}}$, we have
\begin{equation}\label{5}
	\frac{{{\eta }_{{{\lambda }_{0}}}}}{{{M}_{{{\lambda }_{0}}}}}\approx \frac{{{\eta }_{{{\lambda }_{i}}}}}{{{M}_{{{\lambda }_{i}}}}}=R.
\end{equation}
Eq. (\ref{4}) can be simplified to
\begin{equation}\label{6}
	{{I_{m}}}\approx R({{V}_{{A}}}+{{V}_{{{\lambda }_{0}}}})+I_{el}.
\end{equation}
Therefore, by measuring the partial light intensity of the modulated laser, the overall variance $V_{M}={{V}_{{A}}}+{{V}_{{{\lambda }_{0}}}}$ of the modulated light fields can be monitored.

Note that the effect of the Trojan horse attack has nothing to do with the specific value of ${{\lambda }_{0}}$ given that the average photon number of the probe light remains unchanged. Without loss of generality, we can choose ${{\lambda }_{0}}={{\lambda }_{i}}$, therefore ${{V}_{{{\lambda }_{0}}}}\text{=}{{V}_{{{\lambda }_{i}}}}$, $V_{M}={{V}_{{A}}}+{{V}_{{{\lambda }_{i}}}}$ and replace the DBS of Eve with a beam splitter of transmittance ${S}_{2}$ as shown in Figure~\ref{Fig_3}b. Eq. (\ref{6}) can be rewritten as
\begin{equation}\label{7}
	{{I_{m}}}=R{{V}_{{M}}}+I_{el}.
\end{equation}
Therefore, the Trojan horse attack of the eavesdropper is equivalent to the increase of the attenuation of the untrusted quantum channel (Figure~\ref{Fig_3}c), i.e. ${{T}_{1}}=T{S}_{2}$ , where $T$ is the original channel transmittance and ${S_{2}}={I_{{\lambda}_{i}}}/{(I_{{\lambda}_{i}}+I_{{\lambda}_{0}})}$. Since the quantum key distribution protocol is information-theoretical secure for untrusted quantum channel, the Trojan horse attack is discoverable and ineffective.

The measurement of the modulated laser intensity is an average of plenty of measurement data in one data block ($>10^6$). In this case, the effect of the electronic noise can be ignored. Eq. (\ref{7}) can be rewritten as
\begin{equation}\label{8}
	\langle{{I_{m}}}\rangle\approx R{{V}_{{M}}}.
\end{equation}

To defeat Eve's Trojan horse attack, we can estimate the channel transmittance and escess noise using the player's and dealer's data, and the monitored $V_M$. 
\begin{equation}\label{9}
	{{T}_{1}}=\frac{{\left\langle {{X}_{A_1}}{{X}_{B_1}^{m}} \right\rangle}^2 }{\eta {V_{A}V_{M} }}=\frac{T{{V}_{A}}}{{{V}_{M}}},
\end{equation}	
\begin{equation}\label{10}
	\begin{aligned}
		\varepsilon& =\frac{{{V}_{B}}-1-{{\upsilon }_{{el}}}}{{{T}_{1}} }-{{V}_{M}},
	\end{aligned}
\end{equation}	
where $V_B={\left\langle {{X}_{B_1}^{m}}^{2} \right\rangle }$ is the variance of the quadratures measured by Bob. $\eta $ and ${{\upsilon }_{{el}}}$ are the efficiency and electronic noise of homodyne detector, respectively.

\subsection{\label{sec:leve22}Security proof of untrusted source intensity fluctuations}
Since the laser source is untrusted, Eve can also perform source intensity fluctuation attacks. Supposes that the dealer plans to prepare a signal pulse with intensity $I_{\lambda_{i}}^{\prime}$, however, he actually prepares a pulse with the intensity of $I_{\lambda_{i}}^{\prime}\left( 1\text{+}\sigma  \right)$ \cite{41}, where $\sigma$ is the intensity fluctuation caused by the instability of the laser source with mean value zero and variance ${{V}_{\sigma }}$. The intensity of the signal pulse received by the player is $I_{\lambda_{i}}^{\prime}\left( 1\text{+}\sigma \text{+}\varphi  \right)$, where $\varphi $ is the intensity fluctuation caused by Eve's intensity fluctuation attack with mean value zero and variance ${{V}_{\varphi }}$. The actual coherent state that encoding the Gaussian random variables $( {{X}_{A_1}},{{P}_{A_1}} )$ of the player is given by
\begin{equation}\label{11}
	\left|{X}_{A_2}+i{P}_{A_2}\right\rangle=\left| \sqrt{\left( 1\text{+}\sigma \text{+}\varphi  \right)}{{X}_{A_1}}\text{+}i\sqrt{\left( 1\text{+}\sigma \text{+}\varphi \right)}{{P}_{A_1}} \right\rangle . 
\end{equation}

To deal with Eve's source intensity fluctuation attack, we added a photodetector at the player's input port to monitor the intensity of each light pulse and the measured signal is 
\begin{equation}\label{12}
	{{I}_{{m}}^{\prime}}={{I}_{{\lambda_{i}}}^{\prime}}+{{I}_{\text{el}}^{\prime}},
\end{equation}
where ${{I}_{\text{el}}^{\prime}}$ is the electronic noise of the detector. The measured intensity fluctuation of the light pulse relative to the average light intensity is expressed as
\begin{equation}\label{13}
	\delta {{I}}_{m}^{\prime}=\delta {{I}_{{\lambda_{i}}}^{\prime}}+\delta {{I}_{el}^{\prime}},
\end{equation}
where $\delta {{I}_{{\lambda_{i}}}^{\prime}}$ and $\delta {I}_{el}^{\prime}$ are the light pulse  fluctuation and electronic noise fluctuation relative to the average light intensity and they satisfy $\delta {{I}_{{\lambda_{i}}}^{\prime}}\text{=}\sigma \text{+}\varphi $, $\left\langle \delta {{I}_{{\lambda_{i}}}^{\prime}} \right\rangle \text{=}0$, $\left\langle {\delta {{I}_{{\lambda_{i}}}^{\prime}}}^{2} \right\rangle \text{=}{{V}_{{\lambda_{i}}}^{\prime}}$, $\left\langle \delta{{I}_{el}^{\prime}} \right\rangle \text{=}0$, and $\left\langle {{ \delta {{I}_{\text{el}}^{\prime}} }^{2}} \right\rangle={{V}_{\text{el}}^{\prime}}$. 
Considering the source intensity fluctuation, the prepared coherent states can be rewritten as
\begin{equation}\label{14}
	\left|{X}_{A_2}+i{P}_{A_2}\right\rangle=\left| \sqrt{\left( 1\text{+}\delta {{I}_{{\lambda_{i}}}^{\prime}}  \right)}{{X}_{A_1}}\text{+}i\sqrt{\left( 1\text{+}\delta {{I}_{{\lambda_{i}}}^{\prime}} \right)}{{P}_{A_1}} \right\rangle. 
\end{equation}

Notice that the fluctuations of the optical pulses cannot be accurately determined due to the electronic noise of the detector. To ensure the security of the protocol, the player revises his data from $( {{X}_{A_1}},{{P}_{A_1}} )$ to
\begin{equation}\label{15}
	\begin{aligned}
		&{X}_{A_3}={\left( 1+\delta {{I}}_{m}^{\prime}+I_{el}^{max} \right)}{{X}_{A_1}},\\
		&{P}_{A_3}={\left( 1+\delta {{I}}_{m}^{\prime}+I_{el}^{max} \right)}{{P}_{A_1}},
	\end{aligned}
\end{equation}
where $I_{el}^{max}$ is the maximum of the electronic noise of the monitoring detector. In this case, the channel loss and excess noise will be overestimated by the dealer and players \cite{41} (see Section S1, Supporting Information for more details).
\begin{equation}\label{16}
	{{T}_{2}}\approx \frac{{{( 8-{{V}_{{\lambda_{i}}}^{\prime}} )}^{2}}TV_A}{{{( 8\text{+}4I_{el}^{\max }-{{V}_{{\lambda_{i}}}^{\prime}}-{{V}_{el}^{\prime}} )}^{2}}V_{M}},
\end{equation}
\begin{equation}\label{17}
	{{\varepsilon_{1} }}\approx {\frac{( V_B-1-{{\upsilon }_{el}} )}{T_{2} }}-{\left( 1+I_{el}^{max} \right)}V_{M}.
\end{equation}

The fluctuation variance ${{V}_{{\lambda_{i}}}^{\prime}}$ of the light pulse intensity cannot be directly measured. We can only measure the total variance $\left\langle\delta {{I}_{m}^{\prime}}^{2}\right\rangle ={{V}_{{\lambda_{i}}}^{\prime}}+{{V}_{el}^{\prime}}$ of the fluctuation of the laser and the electronic noise, and then subtract the electronic noise variance ${{V}_{el}^{\prime}}$ to obtain the variance ${{V}_{{\lambda_{i}}}^{\prime}}$ of the fluctuation of the laser.

\subsection{\label{sec:level8}Security proof of untrusted source noises}
\begin{figure}[!htbp]
	\centering
	\includegraphics[width=3.4in]{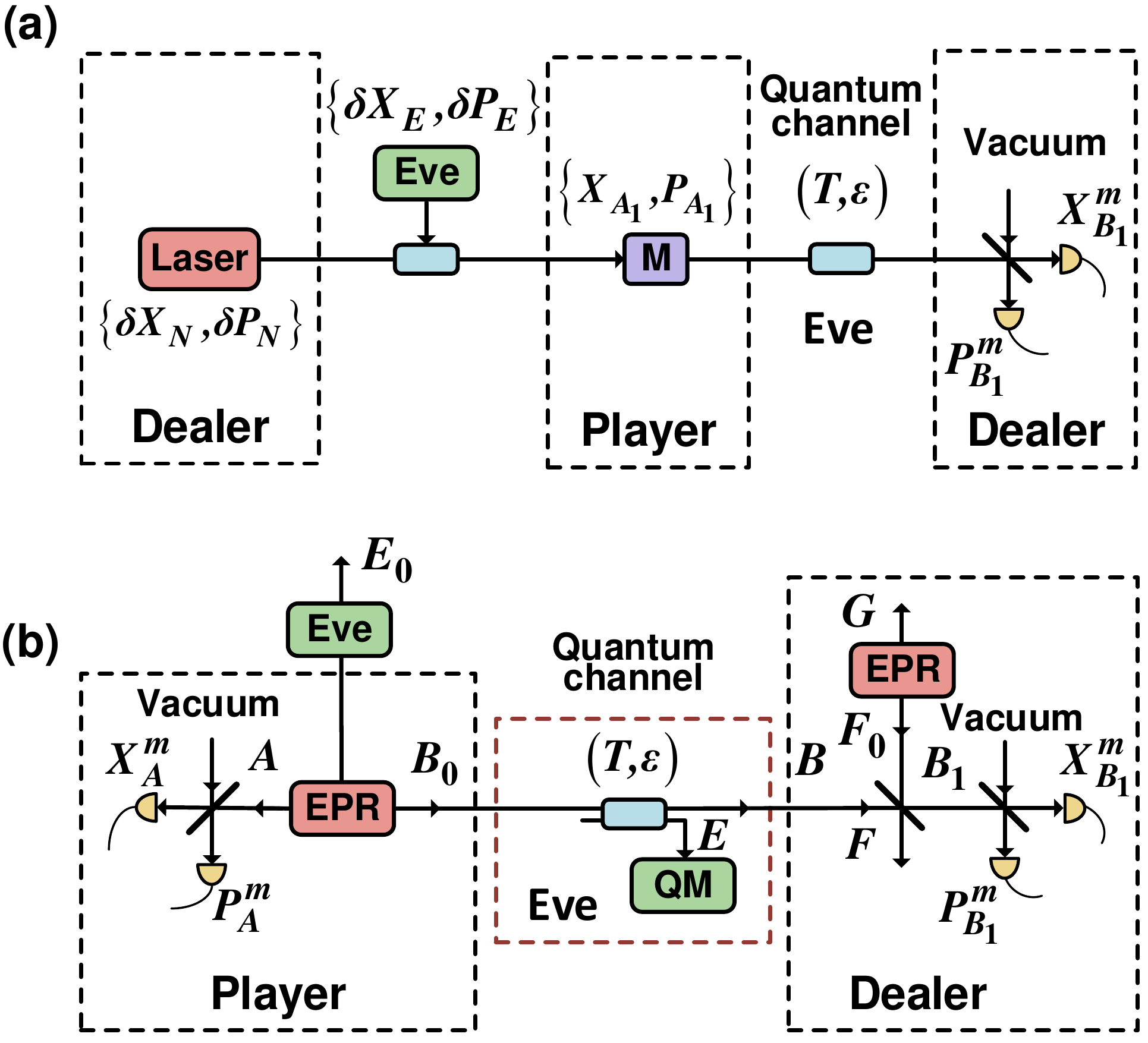}
	\caption{\label{Fig_4}PM and EB schemes of the MSB-QSS and composite QSS protocols with untrusted coherent source. a) The PM scheme. b) The equivalent EB scheme. Eve may introduce noise at the sidebands where the player encoding key information by modulating the laser in the PM scheme. In the equivalent EB scheme, a three-mode entangled state $\rho_{AE_{0}B_{0}}$ is generated with the mode $E_0$ controlled by Eve.}
\end{figure}

In addition to the potential Trojan horse attack and untrusted source intensity fluctuation attack, Eve can also perform source noise attacks. In the following, we will present the Prepare-and-measurement (PM) scheme and the equivalent entanglement-based (EB) scheme.
\subsubsection{\label{sec:level9}PM scheme}
In Figure~\ref{Fig_4}a, a PM scheme is shown. The dealer prepares a coherent state source and its sidemodes quantum state is $\left| {{X}_{N}}\text{+}i{{P}_{N}} \right\rangle$ with $\left\langle \delta {{X}_{N}}^{2} \right\rangle \text{=}\left\langle \delta {{P}_{N}}^{2} \right\rangle =1$ shot noise units (SNU). Eve introduces Gaussian noise $\left\{  \delta {{X}_{E}},\delta {{P}_{E}} \right\}$ on the sidebands where the players encoding the information by modulating the laser, and the noise satisfies  $\left\langle \delta {{X}_{E}}^{2} \right\rangle \text{=}\left\langle \delta {{P}_{E}}^{2} \right\rangle ={{\xi }_{E}}$. The untrusted source received by the player can be expressed as
\begin{equation}\label{18}
	\left| \delta {{X}_{I}}\text{+}i\delta {{P}_{I}} \right\rangle =\left| \delta {{X}_{N}}+\delta {{X}_{E}}+i\left( \delta {{P}_{N}}+\delta {{P}_{E}} \right) \right\rangle. 
\end{equation}

After encoding the key information onto the source, the quantum state of the player is given by
\begin{equation}\label{19}
	\left| {{X_{PM}}}\text{+}i{{P_{PM}}} \right\rangle =\left| \delta {{X}_{I}}+X_{A_1}+i\left( \delta {{P}_{I}}+P_{A_1} \right) \right\rangle.
\end{equation}

The variance of the quadratures for the quantum state is given by
\begin{equation}\label{20}
	\left\langle {{X}_{PM}}^{2} \right\rangle =\left\langle {{P}_{PM}}^{2} \right\rangle=V+{{\xi }_{E}},
\end{equation}
where $V=V_{A}\text{+}1$. The conditional variance of ${X}_{PM}$ (${P}_{PM}$) given $X_{A_1}$ or $\delta {{X}_{E}}$ are
\begin{equation}\label{21}
	{{V}_{\left. {{X}_{PM}} \right|{{X}_{{{A}_{1}}}}}}={{V}_{\left. {{P}_{PM}} \right|{{P}_{{{A}_{1}}}}}}=1+{{\xi }_{E}},
\end{equation}
\begin{equation}\label{22}
	{{V}_{\left. {{X}_{PM}} \right|\delta {{X}_{E}}}}={{V}_{\left. {{P}_{PM}} \right|\delta {{P}_{E}}}}=V.
\end{equation}

\subsubsection{\label{sec:leve20}EB scheme}
The equivalent EB scheme of the MSB-QSS and composite QSS protocol are shown in Figure~\ref{Fig_4}b, a three-mode Gaussian entangled state ${{\rho }_{{A}E_{0}B_{0}}}$ is generated with the mode $E_0$ controlled by Eve. For mode $A$ $\left( X_{A},P_{A} \right)$, mode $E_0$ $(X_{E_0},P_{E_0})$, and mode $B_0$ $\left( {{X}_{B_{0}}},{{P}_{B_{0}}} \right)$, we assume the following realtions are satisfied 
\begin{equation}\label{23}
	\begin{aligned}
		&\left\langle {X_{A}}^{2} \right\rangle =\left\langle P{{_{A}}^{2}} \right\rangle =V,\\
		&\left\langle {{X_{E_0}}^{2} }\right\rangle =\left\langle P{{_{E_0}}^{2}} \right\rangle =1+{{\xi }_{E}},\\
		&\left\langle {{X}_{B_{0}}}^{2} \right\rangle =\left\langle {{P}_{B_{0}}}^{2} \right\rangle =V\text{+}{{\xi }_{E}}.
	\end{aligned}
\end{equation}

The covariance matrix ${{\gamma }_{A{{E}_{0}}{{B}_{0}}}}$ charactering the state ${{\rho }_{A{{E}_{0}}{{B}_{0}}}}$ has the form
\begin{equation}\label{24}
	\left[ \begin{matrix}
		V & 0 & 0 & 0 & \sqrt{{{V}^{2}}-1} & 0 \\
		0 & V & 0 & 0 & 0 & -\sqrt{{{V}^{2}}-1}  \\
		0 & 0 & c{{\xi }_{E}} & 0 & \sqrt{c}{{\xi }_{E}} & 0  \\
		0 & 0 & 0 & c{{\xi }_{E}} & 0 & -\sqrt{c}{{\xi }_{E}}  \\
		\sqrt{{{V}^{2}}-1} & 0 & \sqrt{c}{{\xi }_{E}} & 0 & V+{{\xi }_{E}} & 0  \\
		0 & -\sqrt{{{V}^{2}}-1} & 0 & -\sqrt{c}{{\xi }_{E}} & 0 & V+{{\xi }_{E}}  \\
	\end{matrix} \right]
\end{equation}
where $c\to +\infty $ is a large real number.

The player performs a heterodyne detection on mode $A$ and the measurement result is given by
\begin{equation}\label{25}
	X_{A}^{m}=\frac{1}{\sqrt{2}}\left( {{X}_{A}}+\delta {{X}_{N}} \right),P_{A}^{m}=\frac{1}{\sqrt{2}}\left( {{P}_{A}}-\delta {{P}_{N}} \right).
\end{equation}
The player use the measurement results $\left(X_{A}^{m},{P_{A}^{m}}\right)  $ to estimate the mode $B_{0}$,
\begin{equation}\label{26}
	\begin{aligned}
		&X_{B_{0}}^{\prime}=\frac{\left\langle {{X}_{B_{0}}}{{X}_{A}^{m}} \right\rangle }{\left\langle {{X}_{A}^{m}}^{2} \right\rangle }{{X}_{A}^{m}}=\sqrt{\frac{2(V-1)}{V+1}}{{X}_{A}^{m}},\\
		&P_{B_{0}}^{\prime}=\frac{\left\langle {{P}_{B_{0}}}{{P}_{A}^{m}} \right\rangle }{\left\langle {{P}_{A}^{m}}^{2} \right\rangle }{{P}_{A}^{m}}=-\sqrt{\frac{2(V-1)}{V+1}}{{P}_{A}^{m}}.
	\end{aligned}
\end{equation}

From Eq. (\ref{26}), we have
\begin{equation}\label{27}
	\left\langle X{{_{B_{0}}^{\prime}}^{2}} \right\rangle =\left\langle P{{_{B_{0}}^{\prime}}^{2}} \right\rangle =V_{A}.
\end{equation}

The conditional variances can be expressed as
\begin{equation}\label{28}
	{{V}_{\left. {{X}_{B_{0}}} \right|X_{B_{0}}^{\prime}}}={{V}_{\left. {{P}_{B_{0}}} \right|P_{B_{0}}^{\prime}}}=\left\langle {{X}_{B_{0}}}^{2} \right\rangle -\frac{{{\left\langle {{X}_{B_{0}}}X_{B_{0}}^{\prime} \right\rangle }^{2}}}{\left\langle X{{_{B_{0}}^{\prime}}^{2}} \right\rangle }=1+{{\xi }_{E}},
\end{equation}
\begin{equation}\label{29}
	{{V}_{\left. {{X}_{{{B}_{0}}}} \right|{{X}_{{{E}_{0}}}}}}={{V}_{\left. {{P}_{{{B}_{0}}}} \right|{{P}_{{{E}_{0}}}}}}=\left\langle {{X}_{{{B}_{0}}}}^{2} \right\rangle -\frac{{{\left\langle {{X}_{{{B}_{0}}}}{{X}_{{{E}_{0}}}} \right\rangle }^{2}}}{\left\langle {{X}_{{{E}_{0}}}}^{2} \right\rangle }=V.
\end{equation}

From Eqs. (\ref{26}) and (\ref{27}), the mode $B_{0}$ is projected onto states with variable mean values of (${{{{X}}}_{{{B}_{0}}}^{\prime}}$,${{{{P}}}_{{{B}_{0}}}^{\prime}}$) and corresponding variance of ${{{V}}_{A}}$ conditioned on the player’s measurement. The uncertainty on the inferred values of mode $B_{0}$ for the player (Eq. (\ref{28})) coincides with the noisy coherent state in the PM scheme (Eq. (\ref{21})). Furthermore, from Eq. (\ref{29}), the uncertainty on the inferred values of mode $B_{0}$ for Eve is identical to that in the PM scheme (Eq. (\ref{22})). Therefore, the EB scheme is equivalent to the PM scheme.

\subsection{\label{sec:leve23} Secret key rate}
On the basis of the previous practical security analysis of MSB-QSS and composite QSS scheme, we derive the secret key rate in this part.

The lower bound of the asymptotic secret key rate of the QSS and QCKA protocols against collective attack are given by \cite{42}.
\begin{equation}\label{30}
	K=\beta {{I}_{AB}}-{\chi }_{{BE}},
\end{equation}
where $\beta$ is the reconciliation efficiency, ${{I}_{AB}}$ is the Shannon mutual information between the player and dealer, and ${{\chi }_{BE}}$ is the maximum information available to the dishonest players and eavesdroppers conditioned on dealer’s measurement.

The channel added noise referred to the channel input is given by
\begin{equation}\label{31}
	{{\chi }_{line}}=\frac{1}{{{T}_{2}}}-1+{{\varepsilon }_{1}},
\end{equation}
where ${1}/{{{T}_{2}}}-1$ is introduced by the quantum channel loss. The detection noise referred to the dealer’s input is expressed by
\begin{equation}\label{32}
	{{\chi }_{het}}=\frac{\left[ 1\text{+}\left( 1-\eta  \right)+2{{\upsilon }_{el}} \right]}{\eta }.
\end{equation}

The total noise referred to the channel input is
\begin{equation}\label{33}
	{{\chi }_{tot}}={{\chi }_{line}}+\frac{{{\chi }_{het}}}{{{T}_{2}}}.
\end{equation}
\indent The mutual information ${{I}_{AB}}$ is calculated directly from the dealer’s measured quadratures variance ${{V}_{B}}={{T}_{2}}\eta \left( V\text{+}{{\xi }_{E}}\text{+}{{\chi }_{tot}} \right)$, and the conditional variance ${{V}_{\left. B \right|A}}={{T}_{2}}\eta \left( 1+{{\xi }_{E}}+{{\chi }_{tot}} \right)$
\begin{equation}\label{34}
	{{I}_{AB}}={{\log }_{2}}\frac{{{V}_{B}}}{{{V}_{\left. B \right|A}}}={{\log }_{2}}\frac{\left( V\text{+}{{\xi }_{E}}\text{+}{{\chi }_{tot}} \right)}{\left( 1+{{\xi }_{E}}+{{\chi }_{tot}} \right)}.
\end{equation}

Eve's access information is up bounded by the Holevo quantity
\begin{equation}\label{35}
	{{\chi }_{BE}}=S\left( {{\rho }_{{E_0}E}} \right)-\int{d{{m}_{B_{1}}}p\left( {{m}_{B_{1}}} \right)}S\left( \rho _{{E_0}E}^{{m}_{B_{1}}} \right),
\end{equation}

\noindent where $m_{B_{1}}$ is the measurement of dealer. $p(m_{B_{1}})$ is the probability density of the dealer's measurement outcomes. $\rho_{E}^{{{m}_{B_{1}}}}$ is the quantum state of Eve and dishonest players conditioned on the dealer's measurement result. $S(.) $ denotes the von Neumann entropy. To calculate Eve's accessible information, we know that Eve’s system can purifiy the system ${A{E}_{0}}B$ (Figure~\ref{Fig_4}b), $S\left( {{\rho }_{{{E}_{0}}E}} \right)\text{=}S\left( {{\rho }_{AB}} \right)$, and the system $A{E_0}EFG$ is pure after the dealer's heterodyne measurement, so that $S\left( \rho _{{{E}_{0}}E}^{{{m}_{B_{1}}}} \right)\text{=}S\left( \overset{{{m}_{B_{1}}}}{\mathop{{{\rho }_{AFG}}}}\, \right)$, where $S\left( \overset{{{m}_{B_{1}}}}{\mathop{{{\rho }_{AFG}}}}\, \right)$ is independent of ${m}_{B_{1}}$ for the Gaussian modulated Gaussian states protocol. Now, Eq. (\ref{35}) can be rewritten as
\begin{equation}\label{36}
	{{\chi }_{BE}}=S\left( {{\rho }_{AB}} \right)-S\left( \overset{{{m}_{B_{1}}}}{\mathop{{{\rho }_{AFG}}}}\, \right).
\end{equation}

The covariance matrix of the Gaussian state ${\rho }_{AB}$
\begin{equation}\label{37}
	\begin{aligned}
		{{\gamma }_{AB}}&=\left[ \begin{matrix}
			V{I}  & \sqrt{{{T}_{2}}\left( {{V}^{2}}-1 \right)}{{\sigma }_{z}}  \\
			\sqrt{{{T}_{2}}\left( {{V}^{2}}-1 \right)}{{\sigma }_{z}} & {{T}_{2}}\left( V+{{\xi }_{E}}\text{+}{{\chi }_{line}} \right){I}   \\
		\end{matrix} \right],
	\end{aligned}
\end{equation}\\
where ${I} \text{=}\left[ \begin{matrix}
	1 & 0  \\
	0 & 1  \\
\end{matrix} \right]$ and ${{\sigma }_{z}}=\left[ \begin{matrix}
	1 & 0  \\
	0 & -1  \\
\end{matrix} \right]$.

The symplectic eigenvalues of  ${{\gamma }_{AB}}$ have the form
\begin{equation}\label{38}
	\lambda _{1,2}^{2}\text{=}\frac{1}{2}\left[ A\pm \sqrt{{{A}^{2}}-4B} \right],
\end{equation}
where
\begin{equation}\label{39}
	\begin{aligned}
		& A={{V}^{2}}-2{{T}_{2}}\left( {{V}^{2}}-1 \right)+{{{T}_{2}}^{2}}{{\left( V+{{\xi }_{E}}+{{\chi }_{line}} \right)}^{2}}, \\ 
		& B={{{{T}_{2}}^{2}}}{{\left[ V\left( {{\xi }_{E}}+{{\chi }_{line}} \right)+1 \right]}^{2}}.\\ 
	\end{aligned}
\end{equation}

The symplectic eigenvalues of  $\overset{{{m}_{B_{1}}}}{\mathop{{{\gamma }_{AFG}}}}\,$ have the form
\begin{equation}\label{40}
	\lambda _{3,4}^{2}\text{=}\frac{1}{2}\left[ C\pm \sqrt{{{C}^{2}}-4D} \right],{{\lambda }_{5}}=1,
\end{equation}
where
\begin{equation}\label{41}
	\begin{aligned}
		& C\text{=}\frac{1}{{{\left[ {{T}_{2}}\left( V+{{\xi }_{E}}+{{\chi }_{tot}} \right) \right]}^{2}}} \big\{ A{{\chi }_{het}}^{2}+B+1+2{{\chi }_{het}}  \\ 
		& \times [ V\sqrt{B}+{{T}_{2}}( V+{{\xi }_{E}}+{{\chi }_{line}}) ]+2{{T}_{2}}\left( {{V}^{2}}-1 \right) \big\}, \\ 
		& D={{\left( \frac{V+\sqrt{B}{{\chi }_{het}}}{{{T}_{2}}\left( V+{{\xi }_{E}}+{{\chi }_{tot}} \right)} \right)}^{2}}. \\ 		 
	\end{aligned}	
\end{equation}

The Holevo quantity ${{\chi }_{BE}}$ is given by 
\begin{equation}\label{42}
	\begin{aligned}
		{{\chi }_{BE}}&=G\left( \frac{{{\lambda }_{1}}-1}{2} \right)+G\left( \frac{{{\lambda }_{2}}-1}{2} \right)\\
		&-G\left( \frac{{{\lambda }_{3}}-1}{2} \right)-G\left( \frac{{{\lambda }_{4}}-1}{2} \right),
	\end{aligned}
\end{equation}
where $G\left( x \right)=\left( x+1 \right){{\log }_{2}}\left( x+1 \right)-x{{\log }_{2}}x$.

Using Eqs. (\ref{30}), (\ref{34}), and (\ref{38})-(\ref{42}), we can calculate the lower bound of the secret key rate.

\section{ Experimental section \& methods}
\subsection{\label{sec:leve25}Modulation}
In our experiment, the weak modulation technique is adopted to prepare the coherent states at the sideband modes. Before the modulation, the complex amplitude of a single frequency laser has the form
\begin{equation}\label{43}
	\alpha (t)={{\alpha }_{0}}{{e}^{i2\pi {{f}_{0}}t}},
\end{equation}
where ${{\alpha }_{0}}$ and ${{f}_{0}}$ are the amplitude and frequency of the laser. When the laser is weakly modulated at frequency ${{f}_{i}}$, the sidemode of the modulated laser is given by
\begin{equation}\label{44}
	{\alpha }'(t)={{\alpha }_{0}}({{M}_{a}}+i{{M}_{p}})\left[ {{e}^{i2\pi ({{f}_{0}}+{{f}_{i}})t}}+{{e}^{i2\pi ({{f}_{0}}-{{f}_{i}})t}} \right],
\end{equation} 
where ${{M}_{a}}\ll 1\text{ and }{{M}_{p}}\ll 1$ denote the amplitude and phase modulation depths respectively. Because the average photon number of the carrier satisfies ${{\left| {{\alpha }_{0}} \right|}^{2}}\gg 1$, even if a very weak modulation can faithfully prepare a coherent state with mean photon number of a few photons at the sidemode. From Eq. (\ref{44}), the sidemode states can be written as
\begin{figure}[!htbp]
	\centering
	\includegraphics[width=3.4in]{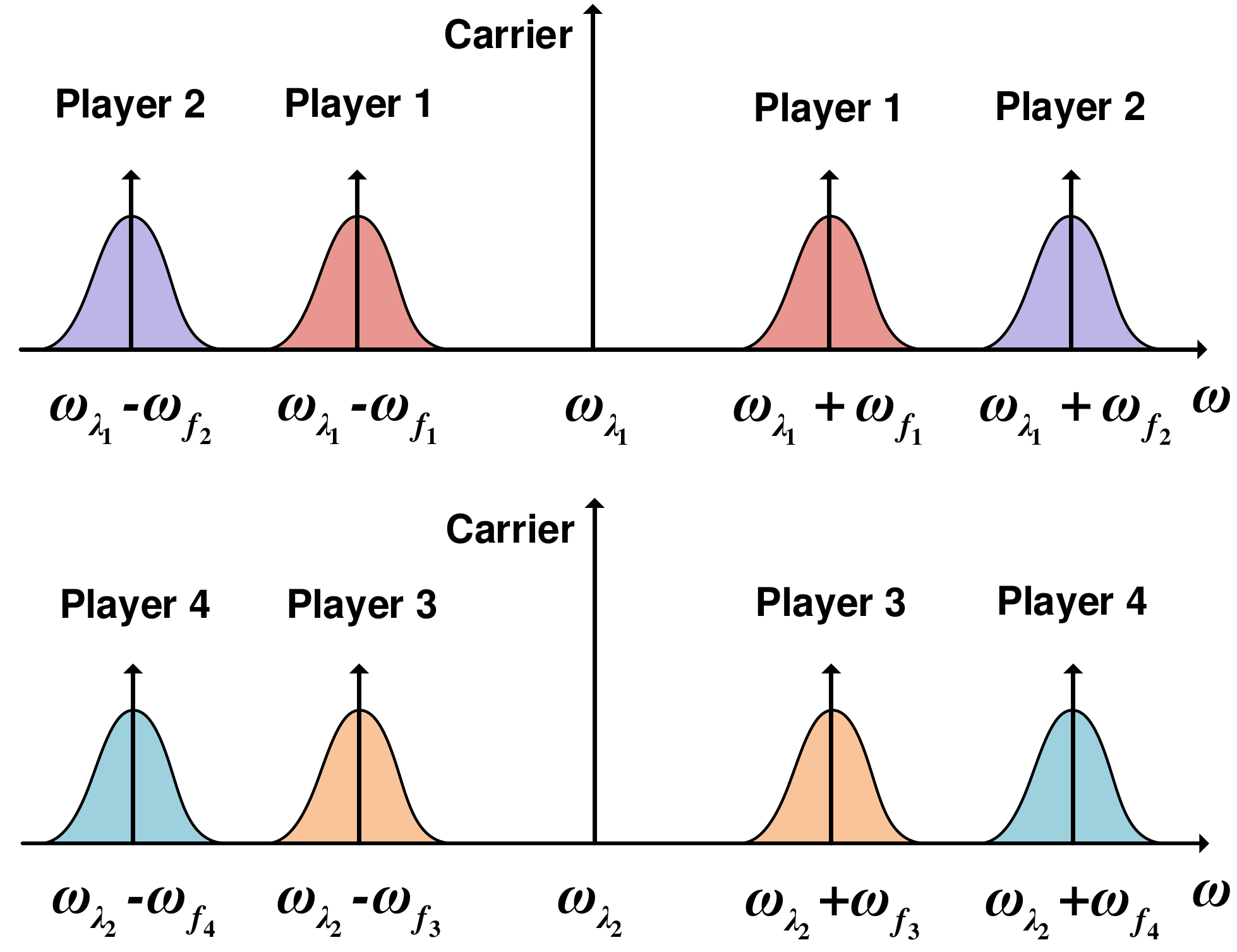}
	\caption{\label{Fig_5}Distribution of quantum signals and carrier signals in frequency domain. The player's signals are generated by modulating the carrier of the lasers, thus the carrier and signals have the same phase. The phase of the signals can be determined by estimating the phase of the carrier.}
\end{figure}

\begin{equation}\label{45}
	\left| {{\phi }_{{{f}_{i}}}} \right\rangle ={{\left| X+iP \right\rangle }_{\pm {{f}_{i}}}}{{\left| 0 \right\rangle }_{f\ne \pm {{f}_{i}}}},
\end{equation}
where $X={{M}_{a}}\alpha_0 $,  $P={{M}_{p}}\alpha_0 $. Therefore, under the condition of large $\left| {{\alpha }_{0}} \right|$ and small modulation depths, we can conveniently prepare sidemode coherent states by weakly modulating the amplitude and phase of the laser field.

\subsection{\label{sec:leve26}Carrier phase evaluation}

In our experiment, the quantum signals and local oscillators (LO) are transmitted through two different long-distance fibers to simulate the local local oscillator (LLO) scheme. In this case, there exists fast phase drifts between the quantum signals and LO. At present, several phase recovered schemes have been proposed that mainly using the pilot-aided feedforward data recovery scheme. The basic idea of the pilot-sequence scheme is to use adjacent pilot pulses to estimate the middle signal’s phase drift \cite{43}. The pilot-multiplexed scheme divides the phase drift into the fast drift and the slow drift parts, and one can implement two remapping procedures to compensate them separately \cite{44}. 

In our scheme, we use a simple method to estimate the phase of the signals. As shown in Figure~\ref{Fig_5}, the quantum signals are generated by modulating the carrier of the lasers and they have fixed phase relations. Although the player's quantum signals are not generated at the same time, the quantum signals and the carrier pass through the same path and are subject to the same phase evolution. Therefore, all the quantum signals have the same phase and we can infer the phase of the quantum signals by evaluating the phase of the carriers.
\begin{figure}[!htbp]
	\centering
	\includegraphics[width=6.8in]{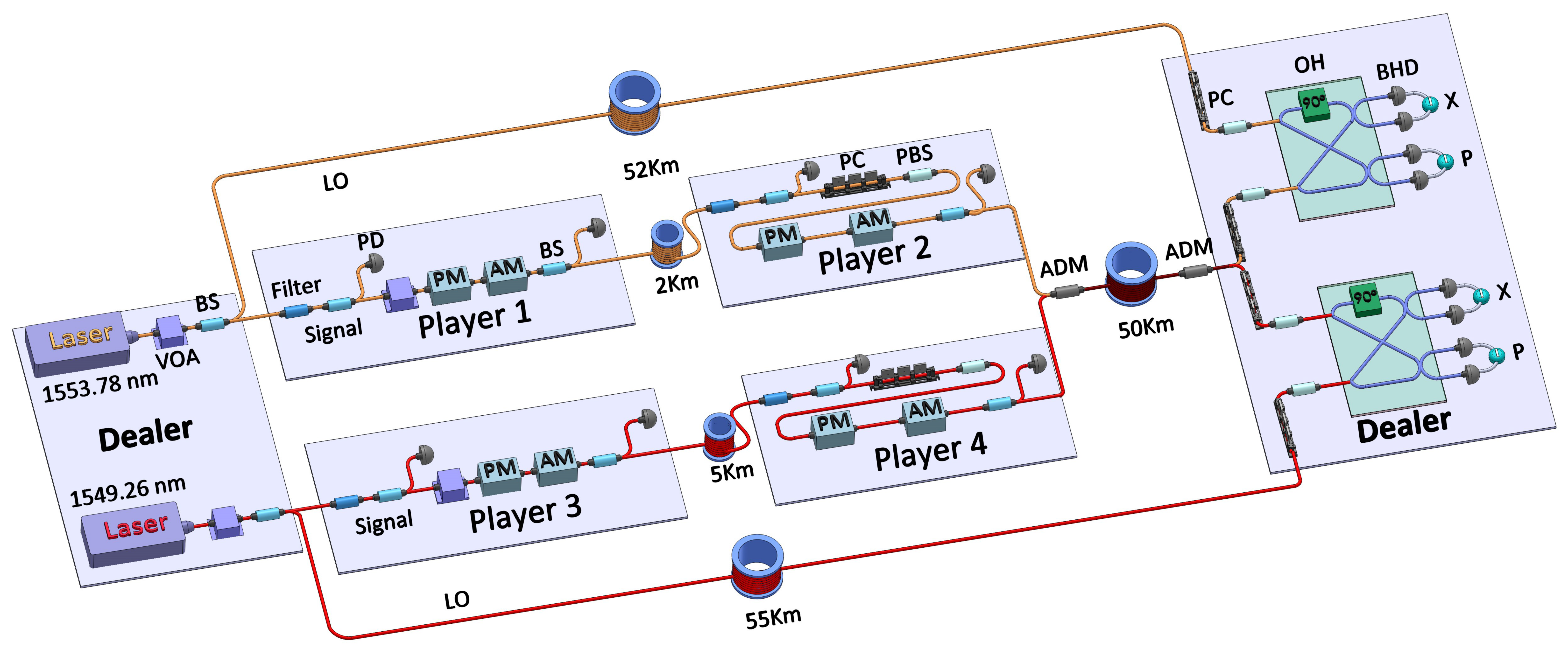}
	\caption{\label{Fig_6}Experimental setup for QSS and QCKA. VOA, variable optical attenuator; BS, beam splitter; PM, phase modulator; AM, amplitude modulator; PD, photoelectric detector; PC, polarization controller; PBS, polarizing beam splitter; OH, optical hybrid; BHD, balanced homodyne detector.}
\end{figure}

\subsection{\label{sec:leve27}Experimental system}

The schematic of the experimental setup is shown in Figure~\ref{Fig_6}. Two continuous wave single-frequency lasers with different wavelengths (1553.78 nm and 1549.26 nm) were prepared by the dealer. A small portion of the lasers are employed as the signals and the rest are acted as the LO fields. The 1553.78 nm and 1549.26 nm signals are sent to the player 1 and player 3, respectively. The players 1 and 3 independently generate two sets of Gaussian random numbers at a repetition rate of 250 kHz, and mix them with a 7 MHz sine signals. Then the mixed signals are loaded on the phase and amplitude waveguide modulators to modulate the two quadrature components of the signal fields. The optical filters at the input port of the players’ station limits the wavelength range of Eve's Trojan horse attacks. To counter against the untrusted source intensity fluctuation attack, a small portion of the incoming signal beams is split and  monitored by a photodetector (PD). The PD after the amplitude modulators (AM) monitors the modulation variance in real time by detecting the intensity of the modulated laser beam. Combine with the optical filter together, they can resist the Trojan horse attacks. 

The modulated signal beams are sent to the players 2 and 4 through a 2 km and 5 km single mode fiber (SMF-28e), respectively. After correcting the state of the polarization by the polarization controller (PC), the players 2 and 4 encode their secret key information at the 9 MHz sideband of the signal beams. By using the ADM, two signal beams are coupled into a 50 km single-mode fiber. The two LO beams are sent to the dealer though 52 km and 55 km single mode fiber, respectively. At the dealer’s station, the signal beams are decoupled by the ADM and measured by heterodyne detection. To this end, two 90° optical hybrid (Kylia) and four balanced homodyne detectors are employed to measure both the amplitude and phase quadrature of the incoming signals. The outputs of the detectors are mixed with 7 MHz and 9 MHz sine waveforms, respectively, and then filtered by two 500 kHz low pass filters. The dealer identifies and extracts the key information of each player in terms of the corresponding wavelengths and sidebands on which the players encoding the information.

\section{Experimental results}
\begin{table}
	\caption{\label{table_1}The parameters of the QSS and QCKA experimental system. $L$, single mode fiber length; $V_{M}$, the overall modulation variance; ${{V}_{el}^{\prime}}$, the electronic noise of the photodetector that monitoring light intensity fluctuations; ${{V}_{\lambda_{i}}^{\prime}}$, the variance of light pulse intensity fluctuation; ${{I}_{el}^{max}}$, the maximum electronic noise of the photodetector that monitoring the fluctuations of the light intensity;  $v_{el}$, the electronic noise of the homodyne detector; $\eta $, the efficiency of the homodyne detector; $\beta$, reconciliation efficiency; $\varepsilon_{1} $, excess noise; $S$, secret key rate. The lower key rates for players 1 and 3 is mainly because that their quantum signals pass through players 2 and 4, resulting in larger channel losses.}
	\resizebox{\textwidth}{20mm}{
		\begin{tabular}{ccccccccccc}
			\hline\hline
			Players&$L$ (km)&$V_{M}$ (SNU)&${{I}_{el}^{max}}$ &${{V}_{el}^{\prime}}$&${{V}_{{\lambda_{i}}}^{\prime}}$&${{v}_{el}}$ (SNU)&$\eta$ (\%)&$\beta$ (\%)&$\varepsilon_{1}$ (SNU)& $S$(bits/pulse) \\ \hline
			Player 1&22&2.14&$4.70\times{{10}^{-4}}$&$5.03\times{{10}^{-8}}$&$1.67\times{{10}^{-7}}$&0.052& 54 & 95 & 0.010 &0.0075\\
			Player 2&20&2.12&$1.75\times{{10}^{-3}}$&$1.87\times{{10}^{-7}}$&$1.94\times{{10}^{-7}}$&0.087& 54 & 95 & 0.011 &0.077 \\
			Player 3&25&2.18&$5.18\times{{10}^{-4}}$&$4.83\times{{10}^{-8}}$&$1.49\times{{10}^{-7}}$&0.045& 56 & 95 & 0.0079&0.0061\\
			Player 4&20&2.08&$1.97\times{{10}^{-3}}$&$1.98\times{{10}^{-7}}$&$1.74\times{{10}^{-7}}$&0.048& 56 & 95 & 0.0071&0.0824\\
			Player 1&52&2.09&$4.94\times{{10}^{-4}}$&$5.06\times{{10}^{-8}}$&$1.64\times{{10}^{-7}}$&0.26 & 54 & 95 & 0.038 &$7.14\times {{10}^{-4}}$ \\
			Player 2&50&2.13&$1.57\times{{10}^{-3}}$&$1.82\times{{10}^{-7}}$&$1.98\times{{10}^{-7}}$&0.39 & 54 & 95 & 0.029 &0.0089\\
			Player 3&55&2.20&$4.97\times{{10}^{-4}}$&$4.88\times{{10}^{-8}}$&$1.53\times{{10}^{-7}}$&0.19 & 56 & 95 & 0.022 &$9.49\times {{10}^{-4}}$\\
			Player 4&50&2.11&$1.82\times{{10}^{-3}}$&$1.89\times{{10}^{-7}}$&$1.83\times{{10}^{-7}}$&0.24 & 56 & 95 & 0.0086&0.013 \\  \hline
		\end{tabular}}
\end{table}
\begin{figure}[!htbp]
	\centering
	\includegraphics[width=6.8in]{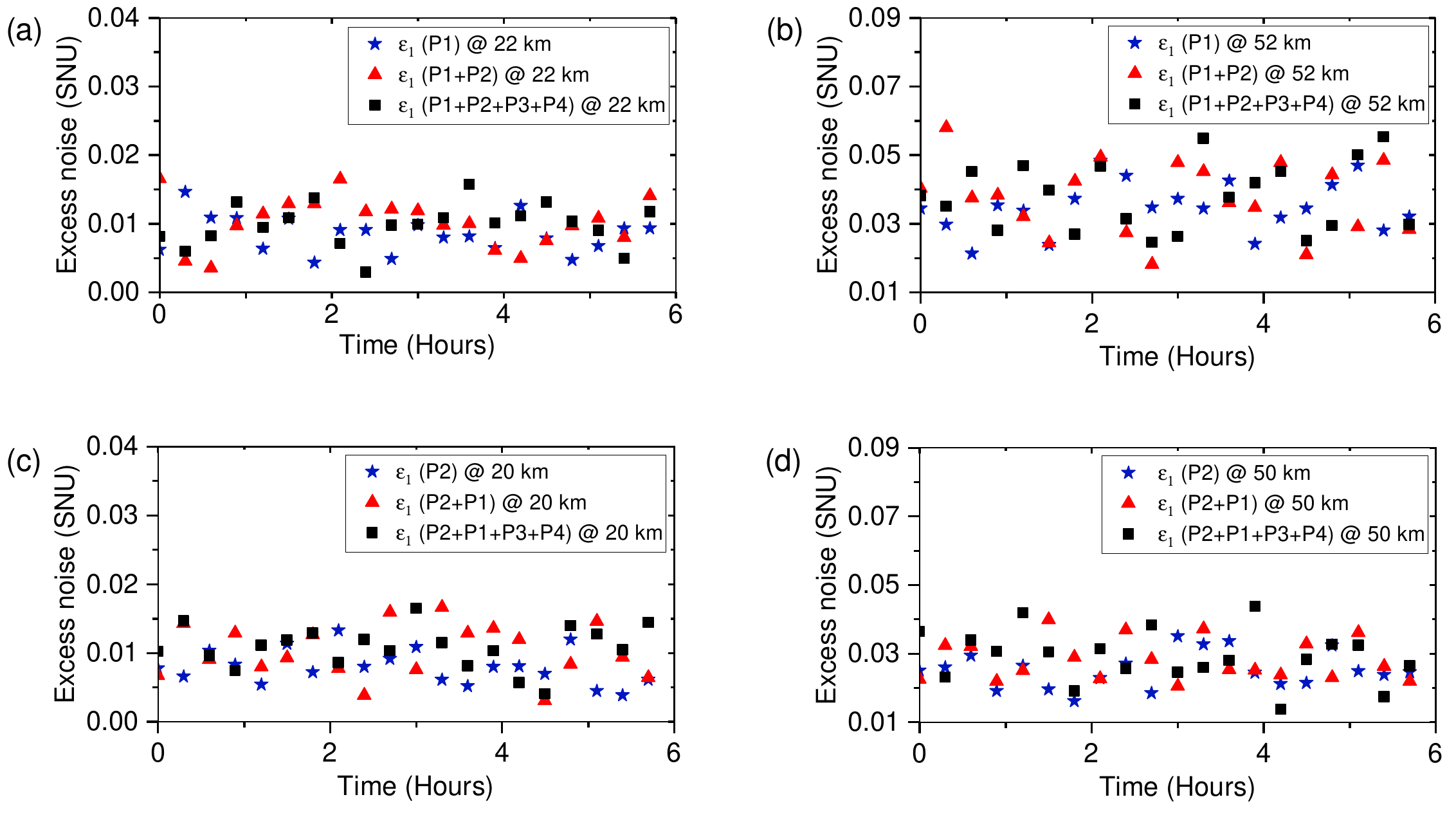}
	\caption{\label{Fig_7} Experimental excess noises of player 1 and player 2 measured over 6 h. a), b) The excess noise of player 1 under different multiplexing methods for total transmission distance of 22 km and 52 km. The blue pentagrams are excess noise with only the player 1 encoding the information. The red triangles represent the excess noise of player 1 when players 1 and 2 encode information at different sidebands. The black squares represent the excess noise of player 1 when players 1, 2, 3, and 4 encode information at different sidebands and wavelengths.  c), d) The excess noises of the player 2 under different multiplexing methods for total transmission distance of 20  km and 50 km. The blue pentagrams are excess noise with only player 2 encoding the information. The red triangles represent the excess noise of player 2 when players 2 and 1 encoded information at different sidebands. The black squares represent the excess noise of player 2 when players 2, 1, 3, and 4 encode information at different sidebands and wavelengths.}
\end{figure}

We demonstrated the proof-of-principle experiment of the proposed QSS and QCKA protocols different long-distance fiber links. The experimental parameters are shown in Table~\ref{table_1}. To investigate the effect of different multiplexing methods on the player's excess noise, we measured the excess noises under different scenarios, only a single player, two players with sideband multiplexing, and four players with both the DWDM and sideband multiplexing. The results is shown in Figure~\ref{Fig_7}. 

In Figure~\ref{Fig_7}a, the average values of the excess noise of player 1 at total transmission distance of 22 km under three cases are 0.00847 (only player 1 encoding the information), 0.0102 (both player 1 and player 2 encoding the information), and 0.098 (all players (1-4) encoding the information), respectively. We can see that the frequency multiplexing has a slight influence on the excess noise. It is possible that the frequency multiplexing causes a little crosstalk during the modulation and demodulation of the quantum signals. For the DWDM of the quantum signals, the player's excess noise has negligible impact on each other. Above phenomenon is also confirmed by the similar results observed in Figure~\ref{Fig_7}b, c, and d.

\begin{figure}
	\centering
	\includegraphics[width=0.4\linewidth]{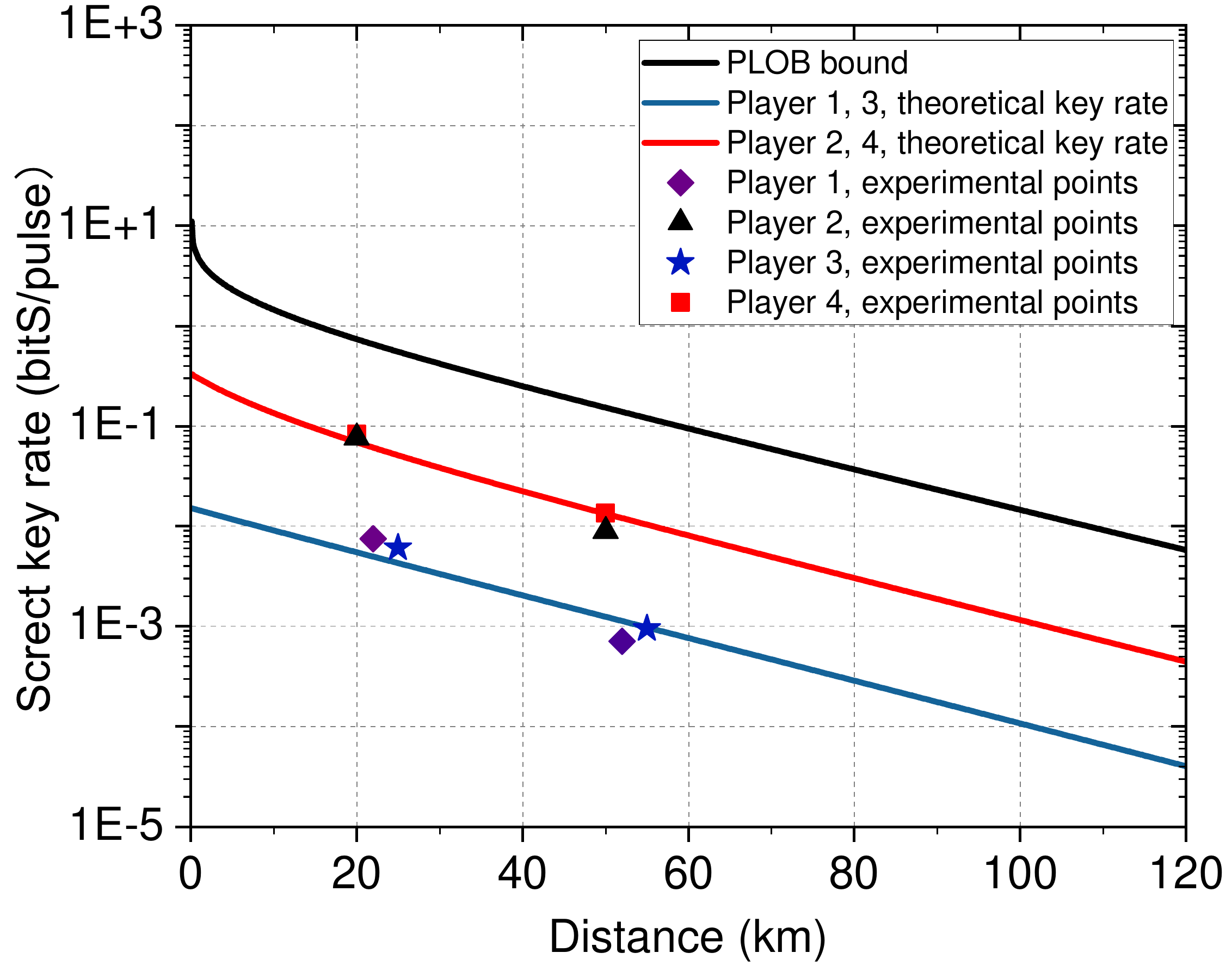}
	\caption{\label{Fig_8} Secret key rates of QSS (QCKA). The black line is the PLOB bound. The two purple rhombus, black triangles, blue pentagrams, and red squares corresponds to the secret key rate of the players 1, 2, 3 and 4 at single mode fiber links of (22 km, 52 km ), (20 km, 50 km), (25 km, 55 km), and (20 km, 50 km), respectively. The blue and red curves represent the simulated secret key rates for the players 1, 3 and the players 2, 4, respectively. The key rate of the QSS (QCKA) at 25 and 55 km fiber links are $0.0061$ and $7.14\times10^{-4}$ bits per pulse, respectively, which are determined by the lowest key rate of the players at 25 km link (player 3) and 55 km (player 1).}
\end{figure}

The experimental secret key rates of the QSS (QCKA) system are shown in Figure~\ref{Fig_8}. The black line represents the Pirandola-Laurenza–Ottaviani–Banchi (PLOB) bound \cite{45}. The two purple rhombus, black triangles, blue pentagrams, and red squares corresponds to the secret key rate of the players 1, 2, 3, and 4 at single mode fiber links of (22 km, 52 km ), (20 km, 50 km), (25 km, 55 km), and (20 km, 50 km), respectively. The blue and red curves represent the simulated secret key rates for the players 1, 3 and the players 2, 4, respectively. Due to the channel loss of the players 1 and 3 is larger than that of the players 2 and 4 (the players 2 and 4 are regarded as eavesdroppers from the viewpoint of the player of 1 and 3), the key rate of the players 1 and 3 are lower. After all players estimate their key rate with the dealer, the lowest  key rate of all the players is set as the key rate for the QSS (QCKA) system. In our case, the key rate of the QSS (QCKA) at 25 and 55 km fiber links are $0.0061 $ and $7.14\times10^{-4} $ bits per pulse, respectively, which are determined by the key rate of players 3 and 1.  
\begin{table}
	\caption{\label{table_2}The insertion loss of the optical devices of the network system. Filter, fiber optical filter; PC, fiber polarization controller; 1/99 BS, free-space 1/99 beam splitter; PBS, free-space polarization beam splitter; PM, free-space phase modulator; AM, free-space amplitude modulator biased at 96\% transmission point; Coupling efficiency, fiber to free-space and free-space to fiber coupling efficiency; ADM, add/drop multiplexer; AWG 48-CH, arrayed waveguide grating of 48-channel.}
	\resizebox{\textwidth}{5.5mm}{
		\begin{tabular}{cccccccccccc}
			\hline\hline
			Equipment components&Filter&PC&1/99 BS&PBS&PM&AM&1/99 BS&Coupling efficiency&ADM&AWG 48-CH \\ \hline
			
			Insertion loss (dB)&0.35&0.1&0.05&0.1&0.05&0.25&0.05&0.2&0.35&3.5\\  \hline
	\end{tabular}}
\end{table}
\begin{figure}[!htbp]
	\centering
	\includegraphics[width=6.8in]{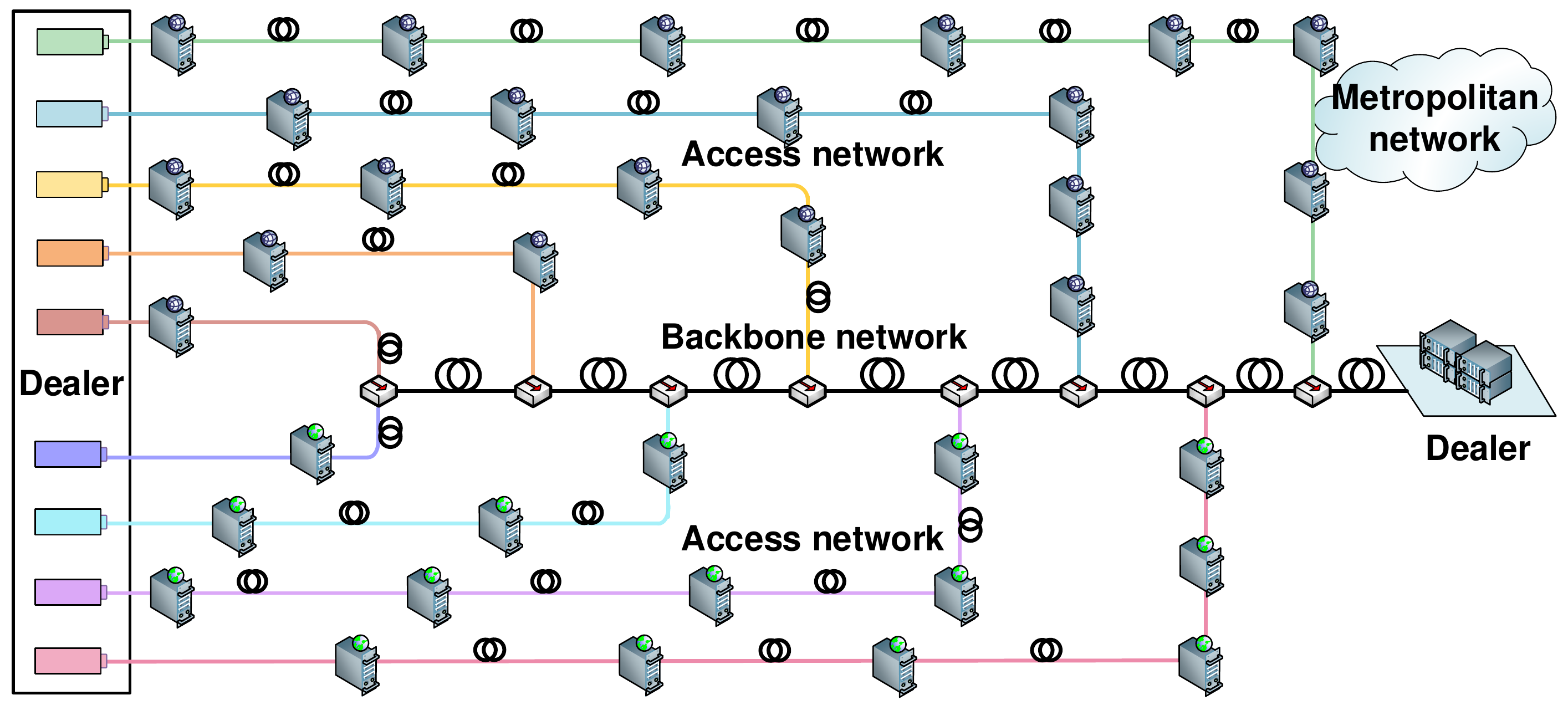}
	\caption{\label{Fig_9}The schematic diagram of the QSS and QCKA network topology. The metropolitan network is composed of the access network and the backbone network. If the total distance between the farthest player in each access network and the dealer is determined, the distance between each adjacent player can be arbitrary, and the distance between the access nodes of the backbone network can be arbitrary.}
\end{figure} 

\section{\label{sec:leve29} Discussion}

On the basis of our presented scheme, in this section we propose the possible construction of a network topology for metropolitan QSS and QCKA network. 

In our proof of principle experiment, we employ the fiber-based components such as amplitude and phase modulators, optical filter, beam-splitters. These fiber pigtailed components have relative large insertion losses, which are detrimental to the performance of the QSS and QCKA protocols. In fact, the players can employ free space optical devices at their stations, which can significantly reduce the adverse insertion losses and increase the player amount.

Table~\ref{table_2} shows the typical losses of state of the art optical devices that are required in the proposed protocol. From the loss values, we can estimate the total insertion losses ($d$) for each player’s station is around 1.35 dB. Using the estimated loss, we propose a possible network topology for metropolitan QSS and QCKA network, as shown in Figure~\ref{Fig_9}. The metropolitan network consists of a backbone network and multiple access networks. The backbone network employs the DWDM-QSS scheme and the access network utilizes the MSB-QSS scheme. The backbone network have $m$ access points, which enables the end players to connect to the network. The upper limit value of the channel loss for each access network is assumed to be $D$. The construction process of the metropolitan network is as follows.

Step 1. The fiber distance between the farthest player and the dealer in each access network is given as $L$, and the channel linear loss is $0.2L$ for standard single mode fiber.

Step 2. Since the insertion loss of the player's station is greater than that of the ADM, in order to maximize the number of the players in the metropolitan network, the backbone network should configure the access nodes as many as possible. When the number of the players in the farthest access network is the least, the number of access nodes of the backbone network are the most. The number of access nodes can be given by
\begin{equation}\label{46}
	m=\frac{D-0.2L-AWG}{ADM},
\end{equation}
where $AWG$ and $ADM$ refer to insertion loss of the AWG and ADM.

Step 3. The number of the players in the $j$th access network counting from receiver can be expressed as
\begin{equation}\label{47}
	n_j=\frac{D-0.2L-jADM-AWG}{d}+1,
\end{equation}
where $j\in \left\{ 1,2,\ldots ,m \right\}$.

Step 4. The maximum number of the players in the entire metropolitan network is
\begin{equation}\label{48}
	N=1+\sum\limits_{j=1}^{m}{{{n}_{j}}}.
\end{equation}
\begin{figure}[!htbp]
	\centering
	\includegraphics[width=3.4in]{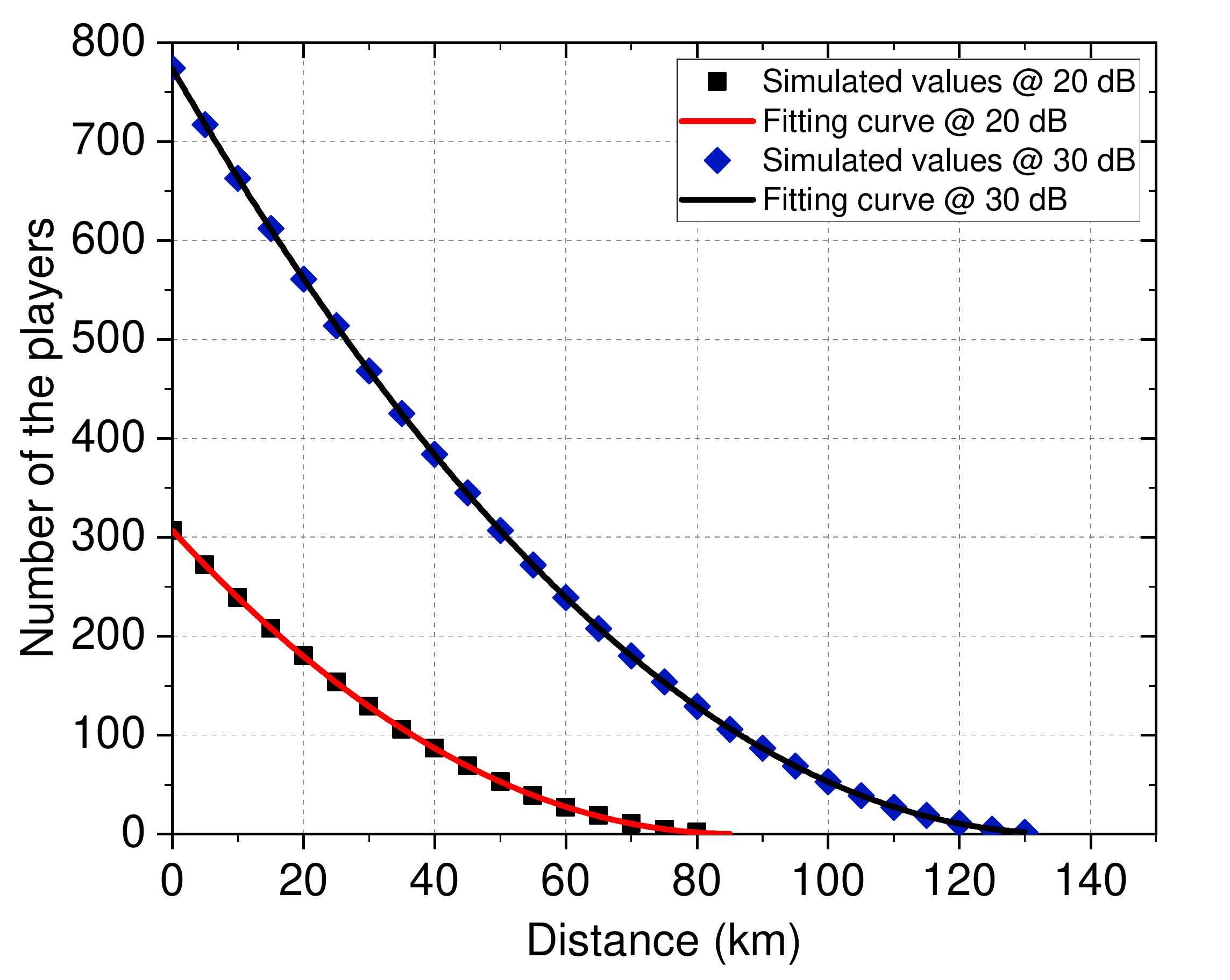}
	\caption{\label{Fig_10} The number of the players versus the transmisssion distance. The black squares and blue rhombus are the simulated number of the players. The red and black curve is the fitting according to the simulated values. Given the channel loss limits of 20 dB and 30 dB, we find that the secure QSS and QCKA among 180 (53) and 651 (307) players are feasible over a metropolitan area over 20 km (50 km).}
\end{figure}  

Notice that, given the fiber distance between the farthest players of the access networks and the dealer, the position of the players in each access network is unrestricted. The span between two adjacent access points can also be freely configured as needed.

Based on the construction method mentioned above, we simulate the maximum number of the players at different transmission distances in Figure~\ref{Fig_10}. The black and blue points represent the results of theoretical simulations under $D=20$ dB and 30 dB. The red and black curves are the fitting curves. The network exhibits a nonlinear dependence between the transmission distance and number of the players. When the channel loss limit is set to 20 dB, we can see that the secure QSS and QCKA among 180 (53) players are feasible within a metropolitan area over 20 km (50 km). If the channel loss limit increases to 30 dB, the number of the players can improve to 651 (307) accordingly.

\section{Conclusion}
In conclusion, we propose three practical, scalable, verifiable $(k,n)$ threshold QSS and QCKA schemes. Our protocols allow each player's secret information to be modulated, transmitted, measured, and processed separately, which effectively eliminate the issues of laser phase locking, excess noise superposition, and preparing laser source for each player. We analyze the practical security for the proposed QSS systems under Trojan horse attack, untrusted sources intensity fluctuating, and noisy untrusted sources. The proposed system are flexible and versatile, they can realize both the QSS and QCKA tasks by just switching the post-processing program. We experimentally investigated the effects of the quantum channel multiplexing of multiple players on the excess noise of the player, and verified the five-party QSS and QCKA quantum communication protocols. A secure key rate of 0.0061 ($7.14\times10^{-4}$) bits per pulse are achieved over 25 (55) km standard single mode fiber. In our protocols, the players do not prepare the lasers and thus the optical module of the player is easy to integrate on a CMOS compatible silicon-based optical chip. Such on-chip integration is beneficial to future applications. Our work provides a feasible solutions for practical quantum private communication network. 

\medskip
\textbf{Supporting Information} \par 
Supporting Information is available from the Wiley Online Library or from the author.

\medskip
\textbf{Acknowledgements} \par 
This work is supported by the National Natural Science Foundation of China
(62175138); Shanxi 1331KSC.

\medskip
\textbf{Conflict of Interest} \par
The authors declare no conflict of interest.

\medskip
\textbf{Data Availability Statement} \par
The data that support the findings of this study are available from the corresponding author upon reasonable request.

\end{document}